# From Neutrino Factory to Muon Collider

*S. Geer*
Fermi National Accelerator Laboratory, P.O. Box 500, Batavia, IL, USA

**Abstract**
Both Muon Colliders and Neutrino Factories require a muon source capable of producing and capturing $O(10^{21})$ muons/year. This paper reviews the similarities and differences between Neutrino Factory and Muon Collider accelerator complexes, the ongoing R&D needed for a Muon Collider that goes beyond Neutrino Factory R&D, and some thoughts about how a Neutrino Factory on the CERN site might eventually be upgraded to a Muon Collider.

## 1    Introduction

Muon Colliders [1] and Neutrino Factories [2] both require a muon source capable of producing and capturing $O(10^{21})$ muons/year within the acceptance of an accelerator. The "front end" required for a Muon Collider (MC) is therefore similar, and perhaps identical, to the corresponding "front end" needed for a Neutrino Factory (NF). Schematics for the two facilities are compared in Fig. 1. The front end muon sources consist of:

–   A high-power multi-GeV proton source with a beam power of ~4 MW or more.

–   A target within a high-field solenoid followed by a $\pi^{\pm}$ decay channel.

–   A system of rf cavities that captures the daughter muons longitudinally into a bunch train, and then applies a time-dependent acceleration that increases the energy of the slower (low energy) bunches and decreases the energy of the faster (high energy) bunches (phase rotation).

–   A cooling channel that reduces the transverse phase space occupied by the beam, so that it fits within the acceptance of the first acceleration stages (NF) or further cooling stages (MC).

Although the schematics for the MC and NF front ends are the same, differences in detail might arise from different requirements on muon bunch structure and repetition rate. To optimize luminosity, a MC will require the muons to be packaged in the minimum number of bunches practical. This is not required for a NF, and hence a NF front end will not necessarily be suitable for a MC. However a MC front end is likely to also be suitable for a NF. Given the strategic value of having an option to upgrade from a NF to a MC, it would seem prudent to keep the upgrade path in mind when designing a NF muon source.

Beyond the front ends there are differences between the two types of facility that are summarized in Table 1. In a NF the beam that exits the front end is accelerated to the energy of choice (a few GeV or a few tens of GeV), and then stored in a ring with long straight sections, at least one of which points towards a distant detector. Muon decays in this straight section produce the desired neutrino beam. In a MC, the beam exiting the front end muon source must be further cooled to reduce the emittance by a large factor, sufficient to achieve high luminosity in the collider. The cooling channel must cool the longitudinal- as well as the transverse-phase-space, and this requires beam cooling technology that goes beyond that needed for a NF. After this 6D-cooling the muon beam is accelerator to the energy of choice (which might be one or a few TeV). Bunches of negative and positive muons are then injected in opposite direction into a storage ring where they circulate and collide at one or more interaction points for about 1000 turns before they decay.

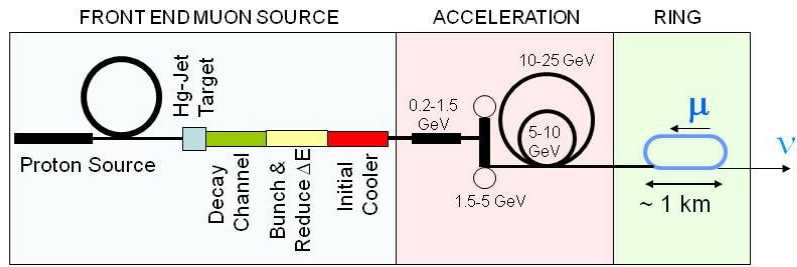

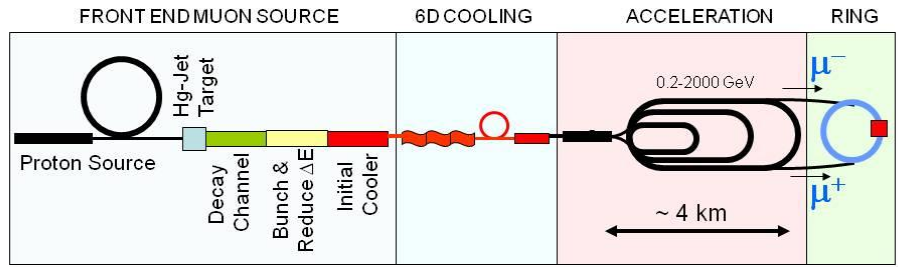

**Fig. 1:** Neutrino Factory and Muon Collider schematics.

**Table 1:** Differences between a Neutrino Factory and a Muon Collider.

| Sub-System | NF | MC |
| --- | --- | --- |
| Cooling | $\varepsilon_\perp \sim 7$ mm | $\varepsilon_\perp \sim 3 - 25$ μm and $\varepsilon_{//} \sim 70$ mm |
| Acceleration | O(GeV) – O(10 GeV) | O(TeV) |
| Storage Ring | Racetrack or triangle | Collider Ring: low β IP, shielding |
| Detector | Magnetized calorimeter | Collider detector |

## 2  Muon Collider Specific Challenges and R&D

Beyond the front end, there are some significant technical challenges for MC designers. In particular (i) achieving a practical design for a cooling channel capable of reducing the 6D beam emittance by a factor O($10^6$) whilst preserving a large fraction of the beam [3], (ii) accelerating the beam to TeV energies rapidly, before the muons decay, in a cost-effective way [4], (iii) achieving low β and hence high luminosity in a ring that is designed to store relatively large emittance beams and cope with backgrounds from muon decay [5], and (iv) shielding the collider detector(s) from the muon-decay induced backgrounds [6].

## 2.1  6D Cooling

In the last three years self-consistent concepts have emerged for a complete MC cooling channel. There are several variants based on different technologies, and aiming at different end-points in 6D phase space. The path that the beam travels in transverse- and longitudinal-emittance-space ($\varepsilon_T, \varepsilon_L$) as it cools has been partially simulated. The result for one candidate scheme is shown in Fig. 2. Note that ionization cooling only reduces the transverse emittance. The longitudinal emittance is reduced by mixing the degrees of freedom as the beam cools. The design of the final cooling stages is particularly challenging. In the present schemes $\varepsilon_L$ is first over-cooled and then allowed to increase as the final ($\varepsilon_x, \varepsilon_y$) reduction takes place.

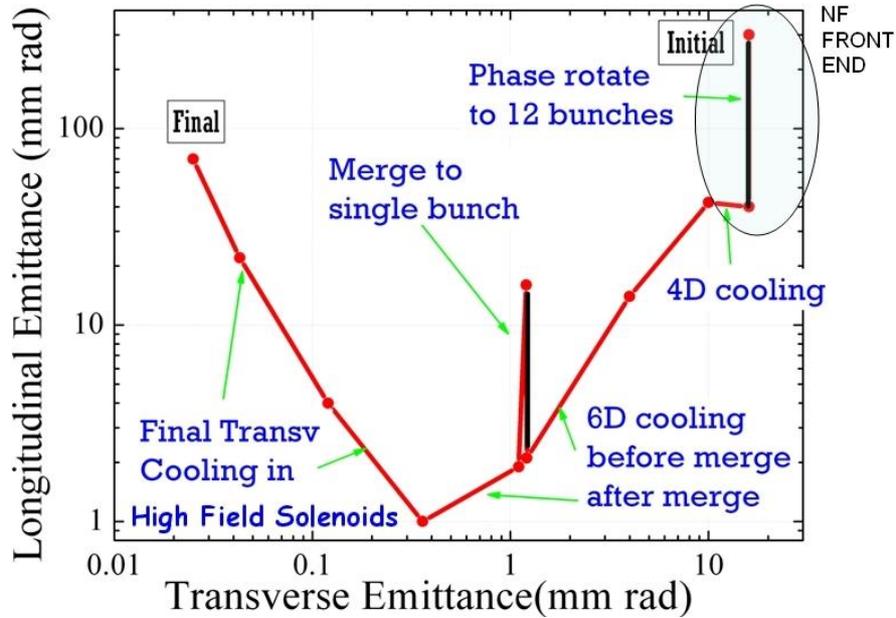

**Fig. 2:** Representative MC 6D cooling scheme from Palmer et al. [7]. The path through 6D phase space from the starting point (top right) to the end point (top left). The scheme begins with a train of 12 bunches from a NF front-end, and then cools the beam by large factors in transverse and longitudinal phase space before combining the bunches (longitudinal emittance jumps by a factor of 12), re-cooling the longitudinal emittance, further cooling the longitudinal and transverse emittance, and finally aggressively cooling the transverse emittance whilst allowing the longitudinal emittance to grow.

All schemes considered so far [3] require components with performances that are beyond the current state-of-art. In particular:

– To continue the battle against scattering which heats the beam as it passes through material in an ionization cooling channel, the solenoid field at the end of the cooling channel must be very high so that the effective focusing angle is much larger than the typical scattering angle. The highest practical field for the last few solenoids has yet to be established, but fields of up to ~50T have been considered. The final MC luminosity is proportional to this field. An R&D program is being pursued to explore the possibility of using High Temperature Superconductor (HTS) solenoids to obtain fields above 30T, and hopefully up to ~50T. Results from conductor studies look promising [8]. It is expected the R&D will also investigate issues associated with how to build HTS solenoids, quench protection etc.

- The mixing of transverse and longitudinal degrees of freedom is accomplished using solenoids "twisted" into a helix, and rf cavities must somehow be integrated into the design. Models consisting of 4 coils have been built and tested [9] to understand the engineering issues involved in building and supporting these quite complicated coil geometries. Once some important design choices have been made for the MC 6D cooling channel, it is planned to build and bench test a 6D cooling channel section, including the RF, coils and energy loss absorber(s).

- RF operation in the appropriate magnetic field configuration must be understood and demonstrated. In most cooling channel designs there are RF cavities operating in a co-axial magnetic field of a few Tesla. It has been shown that when a vacuum copper cavity operating at room temperature is also within a co-axial few Tesla magnetic field its performance is degraded [10]. The maximum accelerating gradient it can support is reduced by typically a factor of two. There are several ideas for how to mitigate this effect. The MUCOOL program, hosted at FNAL, is investigating the options and will characterize the performance that can be assumed for RF cavities operating in a cooling channel.

## 2.2 Acceleration

Acceleration must be rapid to get the muons to high energy before they decay. In principle the early acceleration, up to a few GeV or a few tens of GeV, might be accomplished with the same scheme developed for a NF. Beyond that, acceleration to TeV energies might be accomplished using RLA's (Recirculating Linear Accelerators), or fast ramping synchrotrons. As MC acceleration schemes are further studied, an important consideration is likely to be cost. With respect to using a straight linac for acceleration, you might expect an RLA to have a cost advantage of a factor of ~N, where N is the number of passes the muons make through the accelerating sections. This is not quite true since the cost of the arcs is likely to be significant, and a cost saving of ~N/2 is probably a better approximation. Clearly, it is desirable to maximize N, and perhaps therefore employ fast ramping arcs. Even greater cost savings might be achieved using a fast ramping synchrotron. In fast ramping schemes R&D is needed to develop the required magnets, and in all schemes the SCRF structures must be developed and tested. Initial R&D in both these areas has begun, and needs to pursued in parallel with the MC design studies to inform the choices that must be made in pursuing the ongoing MC feasibility study.

## 2.3 Collider Ring

The final part of the MC facility is the collider ring, where the muon beams collide at low-beta interaction points. The proper design of this ring is a prerequisite for the success of the whole project. There are currently several ring designs under consideration. Recent progress on ring design has been for a 1.5 TeV collider that accommodates a normalized transverse emittance (~25 μm-rad). Other energies, up to 4 TeV, have been studied in the past. The choice of the energy that will be adopted in the coming years for the ongoing MC design feasibility study will be guided by a parallel physics study that is already in progress [11]. The goal of the present MC ring design effort is to develop a lattice design [5] that provides:

- parameters necessary to achieve the design peak luminosity specified by MC physics studies (presently taken as ~ $10^{34}$ cm$^{-2}$ s$^{-1}$),

- momentum acceptance (0.5–1%) and dynamic aperture sufficient to accommodate a muon beam with the emittance expected from the upstream channel,
- reasonable tolerances on field strength, field quality, and alignment errors,
- stability of coherent motion of bunches containing up to $1-2 \times 10^{12}$ muons,
- compatibility with the detector and with protecting the magnets from secondary particles.

The MC ring studies are being pursued in parallel with (i) conceptual studies for collider ring high field dipoles that can operate in the radiation environment created by muon decays at a high energy MC, (ii) detector and background shielding studies since the machine related backgrounds are sensitive to the details of the final focus design.

## 2.4  Detector

A MC detector is imagined to be conceptually similar to detectors at other colliders (vertex detector, tracker, calorimeters, and muon detectors, with the vertex detector and tracker within a solenoid magnet). These detectors must operate in the presence of various backgrounds. Unique to a MC are backgrounds that originate from muon decay. For example, a $\sqrt{s}$=4 TeV MC with one bunch containing $O(10^{12})$ muons would create $O(10^5)$ decay electrons per meter with a mean energy of 700 GeV. This sounds horrendous, but in the late 1990s detailed simulations showed [5] that with a carefully designed final focus system the backgrounds could be reduced so that a MC with luminosity $L\sim10^{34}$ - $10^{35}$ $cm^{-2}s^{-1}$ would have detector background rates comparable to the those at the LHC with $L\sim10^{34}$ $cm^{-2}s^{-1}$. This is possible because the decay electrons born within a few meters of the IP remain within the beampipe in the region of the detector. In the final focus design studied, there was a 130m long straight section on either side of the IP and the last 6.5m was used to shield backgrounds. The shielding occupied cones with cone angles of 20°. Simulations predicted that most of the decay electrons (58%) interact upstream of the shielding, 32% interact in the shielding, and 10% pass through the IP without interacting. As the decay electrons respond to the fields of the final focus system, before they leave the beampipe they lose 20% of their energy by radiating on average 500 synchrotron photons with a mean energy of ~500 MeV. The resulting detector backgrounds have been simulated using two different programs (MARS and GEANT) which yield consistent results. The calculated particle fluxes are used to estimate detector hit rates. For example, consider a cylindrical silicon vertex detector layer at a radius of 10cm. The simulations predict that in $1cm^2$ there with be 750 photons + 110 neutrons + 1.3 charged tracks. These particles yield 2.3 + 0.1 + 1.3 = 4.4 hits $cm^{-2}$. With 300 x 300 $\mu m^2$ pixels this yields an occupancy of 1.3%, which is considered acceptable. The occupancy can be further reduced, by a factor of ~100, by arranging the detector geometry and design so that only coincidences between closely spaced pairs of hits that point back to the IP are read out [18]. In the last few months new detector and background studies have begun, using the latest collider lattice and improvements to the final focus. Preliminary results suggest that the shielding cone angle can be reduced to 10 degrees without significantly increasing the detector background.

## 3  European Contributions

Initial design and physics studies for muon storage ring based accelerator facilities was conducted at CERN in the late 1990's [12]. This has been followed by participation in design, simulation, and experimental studies aimed specifically at Neutrino Factories. The European contributions with perhaps the highest impact have been:

- The HARP experiment [13] at CERN. Detailed measurements of particle production from HARP have been important in constraining and improving the particle production models used to optimize the target design and pion collection scheme. We now have increased confidence in the calculated muon yields for a given target design.

- Hosting the MERIT experiment [14] at CERN. This crucial proof-of-principle liquid mercury target demonstration experiment has shown that this technology can be made to work with the beam parameters assumed for NF and MC front ends.

- The International Scoping Study (ISS) [15]. This study, initiated by the RAL management, was the first NF study that was fully inter-regional in its organization and participation. The ISS prepared the way for the ongoing International Design Study for a NF (IDS-NF).

- The MICE experiment [16] at RAL. This key demonstration of ionization cooling channel technology is expected to be completed in a few years. It will establish the ability of our simulation tools to accurately describe what happens when muons pass through a cooling channel consisting of energy absorbers and RF re-acceleration within a system of solenoids. It will also produce a cadre of engineers and scientists that understand the various aspects of designing, building and operating ionization cooling channels.

- The EMMA experiment [17] at Daresbury. This experiment will study beam dynamics in a new type of accelerator (a so called "non-scaling FFAG") that emerged from work on NF and MC designs, and that is a candidate for a cost-effective "high-energy" acceleration stage for a NF.

- The NUFACT workshops. This series of International workshops was a European initiative launched at Lyon in 1999, following the Fermilab "Workshop on Physics at the First Muon Collider and at the Front End of a Muon Collider", in which the NF concept was proposed, at the end of 1997. The NUFACT workshops are unique in bringing together accelerator and particle physicists focused on future neutrino facilities, and have become an important tool in fostering and discussing relevant international initiatives (examples are MICE and the ISS).

In addition to these high profile activities, there has also been European involvement in the U.S. Neutrino Factory and Muon Collider Collaboration R&D program, and this involvement has made possible many of the communities' international initiatives over the years.

## 4    Outlook

The next steps for NF and MC R&D are:

- IDS-NF: The International Design Study for a Neutrino Factory aims to deliver a Reference Design Report by ~2013. In addition to deepening our understanding of NF design, performance, and cost, it will also take our understanding of the front end needed for a MC to the level appropriate for an RDR.

- MUCOOL and MICE: By ~2013 the U.S. program MUCOOL and the International experiment MICE will have respectively mapped out the performance we can expect for RF cavities operating within a cooling channel and demonstrated the performance of a short cooling section in a muon beam. These are critical steps in building confidence that we know how to design ionization cooling channels and that they will work as expected.

- MAP: The existing MC and NF R&D endeavors in the U.S. are being consolidated into a new national program (Muon Accelerator Program – MAP) hosted at Fermilab. MAP is expected to receive significantly increased support. In addition to participating in the IDS-NF, the main task for MAP is a Design Feasibility Study for a MC on a timescale that is relevant for the next round of decisions that the HEP community must make on collider facilities beyond the LHC.

- MC Physics, Detector and Background Studies: It has been more than 10 years since the last round of MC machine-detector interface studies were performed. The results at that time from detailed shielding deigns and background simulation studies were encouraging, with predicted background hit densities in the various detectors at levels that were considered acceptable provided there was a tungsten shielding cone with an opening angle of 20 degrees in the forward direction. In the last decade there have been significant changes. On the one hand detector technology has greatly advanced and MC lattice designs have made progress. On the other hand, the community expectations for detector performance have become more demanding. With this is mind, a new machine-detector interface study was initiated at a Fermilab workshop in November 2009. Since then, with the latest MC lattice, an improved shielding design, and a full detector simulation, preliminary results suggest the shielding cone angle can be reduced to 10 degrees with backgrounds comparable to those reported in the old early studies. The present machine-detector interface studies are expected to continue until the summer of 2011, and produce a detailed report on MC physics, detectors and backgrounds.

## 5    A Muon Collider at CERN

Multi-TeV Muon Colliders have potential advantages over their multi-TeV electron-positron collider cousins that go beyond a well defined centre-of-mass energy, higher ultimate energy and significantly lower cost. Muon Colliders are also expected to have a much smaller footprint, which would enable a multi-TeV MC to fit within the boundaries of existing accelerator site, and they offer interesting staging possibilities with physics at each stage. This gives flexibility as new physics beyond the Standard Model unfolds. This flexibility is illustrated in the staging options illustrated in Fig. 3.

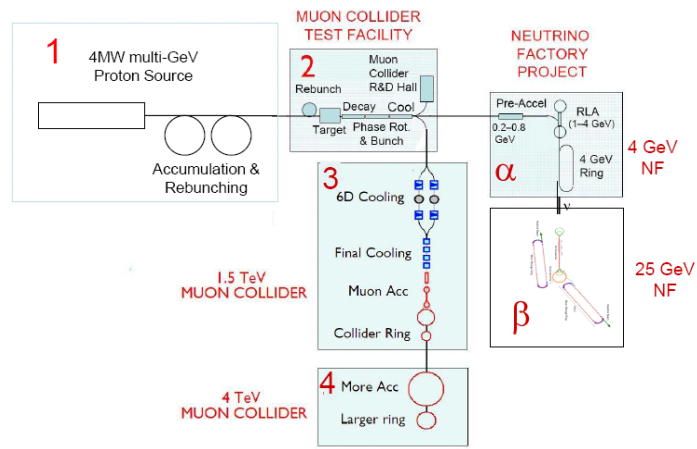

**Fig. 3:** Schematic illustrating the staging possibilities on the path to a multi-TeV Muon Collider.

The footprint of multi-TeV MC would fit on the CERN site, and an upgraded version of the existing CERN proton source could drive the front end muon source needed for a NF and/or a MC. One can imagine the proton source upgraded to produce a 4 MW beam which, together with suitable accumulation and rebunching rings, and a suitable target facility, solenoid decay channel, and muon bunching and phase rotation channel, would provide a unique high-intensity muon source suitable for a suite of low energy muon experiments, whilst also providing a NF/MC test facility.  Following this, or as a part of this first step, there could be a low energy NF, a high energy NF, an initial TeV-scale MC, or the full multi-TeV MC.  Thus, there are many options that can be tailored to emerging physics needs and funding constraints.

## 6    Summary

Progress on MC R&D in the last few years has been encouraging. Although there are tough design and technology challenges to be met before a multi-TeV MC can be proposed, the physics- and cost-incentives to meet these challenges are great. Over the last decade there has been significant European participation in the leadership and execution of the muon accelerator R&D program. A vision has emerged of an accelerator complex based on an intense muon source that could be developed in stages, and lead to an affordable multi-TeV MC that would fit, for example, on the CERN site. The next steps towards this vision are the IDS-NF to deliver a NF RDR, and a MC design feasibility study. In a few years we hope to show that a MC is an option for high energy physics beyond the LHC.

**Acknowledgement**

This work is supported at the Fermi National Accelerator Laboratory, which is operated by the Fermi Research Association, under contract No. DE-AC02-76CH03000 with the U.S. Department of Energy.